\documentclass[11pt]{article}
\usepackage{jheppub}
\usepackage{pstool}
\pdfoutput=1
\usepackage{color}
\usepackage{float}
\usepackage{array}
\usepackage{amssymb}
\usepackage{amsmath}
\usepackage{amsthm}
\usepackage{graphicx}
\usepackage{caption}
\usepackage[labelsep=quad]{subcaption}
\usepackage{epstopdf}
\usepackage{placeins}
	
\usepackage{epsfig}

\def\pb[#1,#2]{\{#1, #2\}}
\def\deb[#1,#2]{[#1,#2]_{\text{D.B.}}}

\def\l{\lambda}

\def\ep{\epsilon}

\def\Or[#1]{{\text{O}}\left({#1}\right)}
\def\dotl[#1,#2]{\left\langle #1,\, #2 \right\rangle}
\def\dotlb[#1,#2]{\left\langle #1,\, #2 \right\rangle}
\def\dotlm[#1,#2]{\left[ #1,\, #2 \right]}
\def\dotp[#1,#2]{(\vect{#1} \cdot\vect{#2})}
\def\aff[#1,#2]{\hat{#1}(#2)}

\def\n4sym{{\cal N}=4 SYM}
\def\>{\rangle}
\def\<{\langle}
\def\weight[#1,#2,#3]{\{(#1),#2,#3\}}
\def\ads[#1]{$\text{AdS}_{#1}$}

\newcommand{\be}{\begin{equation}}
\newcommand{\ee}{\end{equation}}
\newcommand{\ba}{\begin{align}}
\newcommand{\ea}{\end{align}}
\newcommand{\bs}{\begin{split}}
	\def\sess\end{split}
\newcommand{\vect}[1]{{\boldsymbol{#1}}}

\def \bea {\begin{eqnarray}}
\def \eea {\end{eqnarray}}
\def \bea* {\begin{eqnarray*}}
	\def \eea* {\end{eqnarray*}}

\def \be {\begin{equation}}
\def \ee {\end{equation}}
\def \bes {\begin{equation*}}
\def \ees {\end{equation*}}

\newtheorem{theorem}{Theorem}

\def\th{\theta}
\def\l{\lambda}

\def\ep{\epsilon}

\title{Vacuum block thermalization in semi-classical 2d CFT}
\author[a]{Gideon Vos,}
\emailAdd{g.vos@rug.nl}
\affiliation[a]{Van Swinderen Institute for Particle Physics and Gravity, \\ University of Groningen, Nijenborgh 4, 9747 AG, The Netherlands \\}

\date{}

\abstract{The universal nature of black hole collapse in asymptotically $AdS_3$ gravitational theories suggests that its holographic dual process, thermalization, should similarly be fixed by the universal features of 2d CFT with large central charge $c$. It is known that non-equilibrium states with scaling dimensions of order $c$ can be sorted into states that eventually thermalize and those that fail to do so. By proving an equivalence between bounded Virasoro coadjoint orbits and certain (in)stability intervals of Hill's equation it is shown that a state that fails to thermalize can be promoted to a thermalizing state by preparing the system beforehand with an energy greater than an appropriate threshold energy. It is generally a difficult problem to ascertain whether a state will thermalize or not. As partial progress to this problem a set of lower bounds are presented for the threshold energy, which can alternatively be interpreted as criteria for thermalization.}
\keywords{AdS/CFT, Black Holes, Conformal Field Theory, Thermalization}
\begin{document}
	\maketitle

\section{Introduction}	
Strongly coupled quantum field theories provide examples of very interesting but extremely complex chaotic phenomena. Unfortunately the amount of precise statements that can be made about systems at strong coupling is very limited, and these typically only apply to very specific models. One family of strongly coupled theories where one can make rather powerful statements is 2d conformal field theory. In particular, if one narrows it down further and considers 2d CFT in the regime of large central charge, in which case the amount of accessible information grows even further. 

Better yet a lot of these statements are often universal, depending only on conformal symmetry, modular invariance or the presence of the identity operator in the operator spectrum, while otherwise completely independent of the microscopic details of the theory. Examples include the famous Cardy formula \cite{Cardy},\cite{Hartman:2014oaa} and the universal ground state entanglement entropy \cite{Hartman:2013mia}\cite{Chen:2016dfb}. A particularly fruitful common feature of 2d CFT is the universal contribution of the vacuum representation to the Hilbert space and its state-operator dual, the identity operator. In the 't Hooft limit in higher dimensions the presence of the identity operator block ensures large-$N$ factorization, in 2d dimensions the extended vacuum representation under the much larger Virasoro algebra causes the identity block to capture substantially more non-trivial information of the field interactions, in particular this has been exploited in for instance \cite{Fitzpatrick:2015zha}\cite{Anous:2016kss}.


Of course one of the most, if not the most, exciting developments in theoretical physics of the last decades is the AdS/CFT correspondence. Its most pressing contributions are in the form of implications to quantum gravity. Rightfully so, most of the AdS/CFT literature is devoted to formulating (or possibly defining) quantum gravity in terms of conformal field theory. But the flipside of the conjecture provides us with a lot of intuition on strongly coupled conformal field theory with large numbers of degrees of freedom in terms of semi-classical gravity. For one we expect the gravitational sector of the $AdS_3$ bulk dual to exhibit a phase transition in the form of black hole collapse. In the boundary CFT we hence expect that non-equilibrium states evolve at asymptotically late Lorentzian times to either evolve to an equilibrium state that is either in an integrable phase or a thermal phase. 

The key word in the CFT description of black hole collapse is thermalization, it is typically said that a state thermalizes if at asymptotic late times a certain class of sufficiently `simple' correlators on the state approach the expectation values that would have been obtained if they had been computed on a thermal state instead. The subject of this paper will be to consider the identity block contribution to a two-point function of light probe operators on a heavy non-equilibrium state, specifically at leading order in a large central charge expansion.

\subsection*{Summary of the central message}

The distinguishing function determining whether a state thermalizes or not is the stress tensor expectation value of the state. A family of stress tensor expectation values which are connected to each other through conformal transformations form a Virasoro coadjoint orbit, all of these distinct orbits have been classified \cite{Balog:1997zz,Witten,Segal,Lazutkin}. While the role of these orbits within the context of 3d gravity with a negative cosmological constant is known \cite{Banados:1998gg,Martinec:1998wm,Sheikh-Jabbari:2014nya,Compere:2015knw,Sheikh-Jabbari:2016unm,deBoer:2016bov,Balasubramanian:1999re,Cotler:2018zff}, in \cite{Banerjee:2018tut} the authors established from the conformal field theory perspective that the orbit in which the stress tensor expectation value of a state is contained determines its ultimate fate.

In this paper it will be demonstrated that given a non-equilibrium state that would keep the system in an integrable phase, there exists a state associated energy scale such that one can trigger thermalization. Specifically if we prepare the system in advance in an energy eigenstate with an energy above this scale and release the non-equilibrium state on top of this eigenstate the system will thermalize. Furthermore it will be shown that unitarity combined with Virasoro representation theory ensures that there can only be at most one such transition scale per state.

\begin{figure}
	\centering
		\includegraphics[scale=0.2]{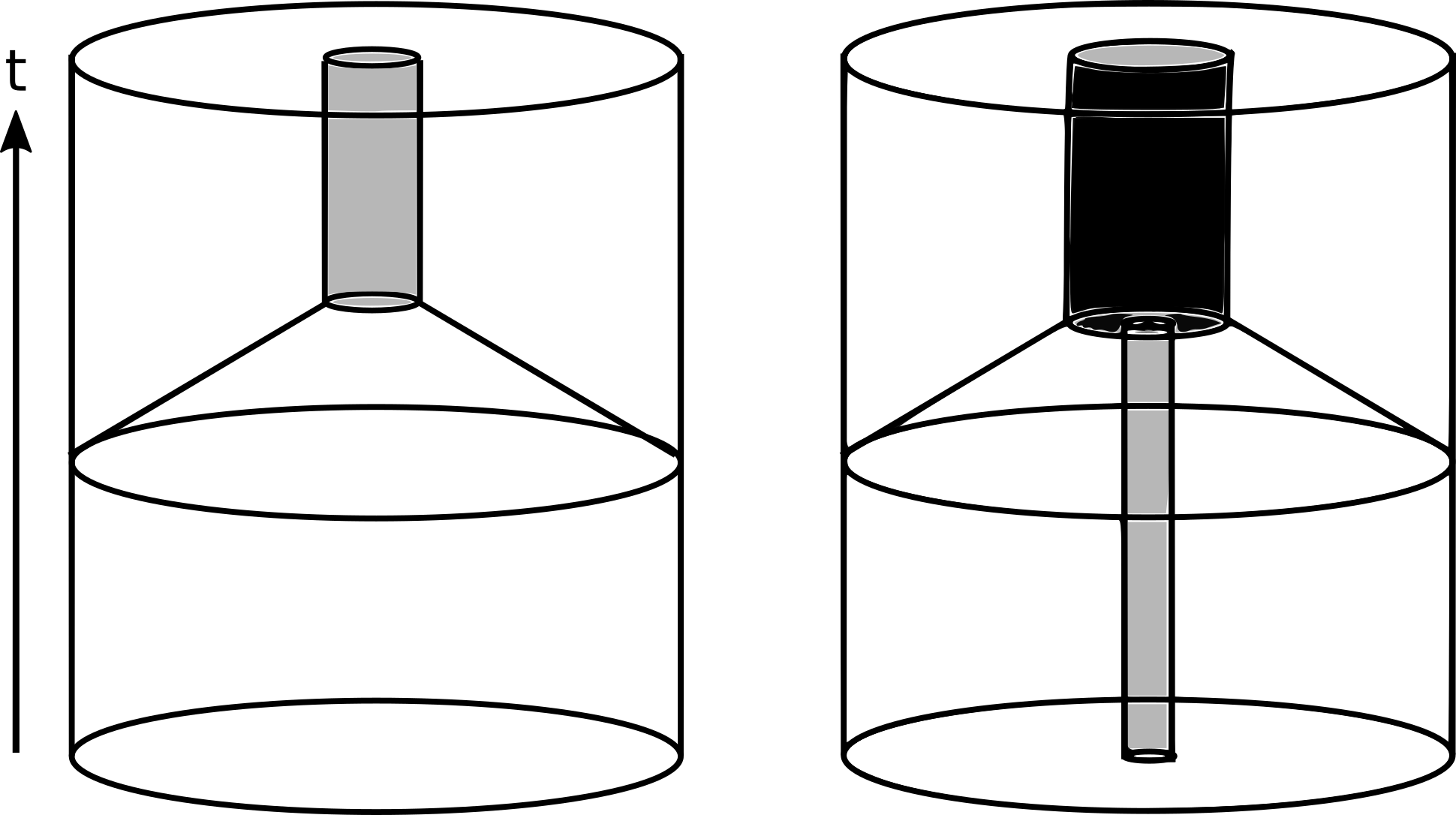}
	\caption{The bulk gravitational picture is particularly intuitive. If a non-equilibrium collapse state is prepared that will collapse into a non-thermal geometry (e.g. a conical defect geometry), then if the experiment were to be repeated with an initial conical defect with mass $m$ in the asymptotic past, then there exists a critical value of $m$ such that the conical defect is pushed over the critical opening angle and a BTZ black hole is formed instead.}
	\label{CollapseDiagram}
\end{figure}

While the gravitational bulk picture of this phenomena is very intuitive (see figure \ref{CollapseDiagram}) a pure conformal field theory description of this transition for CFTs with large numbers of degrees of freedom appears to be lacking in the literature, this paper attempts to fill that gap. 

on a more pragmatic note, this approach naturally presents a set of lower bounds on the transition energy in terms of the other Virasoro conserved charges of the state. These provide us with a set of necessary (but not sufficient) conditions that a state needs to satisfy in order to thermalize. It is an open problem to find an analytically accessible diagnostic deciding whether a state thermalizes or not \cite{deBoer:2016bov}, in principal the answer is given by a path-ordered integral but it is generally not feasible for these to be analytically computed. The necessary conditions in terms of Virasoro charges gives a partial answer to this question in terms of much simpler contour integrals instead. Alternatively, if the Virasoro conserved charges are interpreted as thermodynamic equation of state variables, then the derived lower bounds can be thought of as bounding the equation of state in the thermalizing phase.

This paper has been organized in the following way, section 2 provides a relatively low-technical review of the identity block contribution to probe two-point functions on heavy states and the role of Virasoro coadjoint orbits. Section 3 discusses some of the general features of Hill's equation, the class of ODEs that naturally occur in these two-point functions and presents an equivalence between a subspace of Hill's equations and the space of bounded Virasoro orbits. In section 4 a set of lower bounds are derived for the transition energy from an elliptic to hyperbolic state. Furthermore two appendices have been added, the first of which contains a more technical review of the classification of Virasoro coadjoint orbits. The second appendix provides some of the more technical aspects of the proof for the argued equivalence in section 3. For a reader who is interested in an overview of the results but wishes to avoid the technical derivations reading section 2 is in principal sufficient.

\section{Black hole collapse and the Virasoro identity block}
One particular feature of asymptotically $AdS_3$ gravity is that its black hole solutions, the BTZ black holes, possess a mass gap. As a result not all collapsing states can form black holes. Naively one could expect that the total energy of a state could act as a diagnostic to determine whether a collapse process' final state consists of a black hole. Interestingly this does not quite work due to the rich boundary dynamics attributed to the Brown-Henneaux conserved charges \cite{Brown:1986nw}. Simply put, arbitrarily large amounts of energy can be stored in the Virasoro boundary charges without contributing to a potential final state black hole mass, in fact one could remove or increase this energy by means of a conformal transformation \cite{deBoer:2016bov}. As an analogy, one can consider a bound system of particles, one could increase the energy of this system by arbitrarily large amounts by boosting it to higher and higher velocities but it will not affect the Schwarzschild radius of the bound system.

By the holographic nature of asymptotic $AdS_3$ gravity it is expected that there exists a CFT analogue to this dichotomy between states $|V\rangle$ that collapse to a black hole and states that fail to form black hole states. The holographic dictionary states that the analogue would be that some initial states will equilibriate to states that are indistinguishable from being thermal \cite{srednicki1999approach} while other states remain in an integrable phase. The class of 2d CFTs under consideration will be very broad, mirroring the universality of black hole thermodynamics in gravitational theories. The focus will be on 2d CFTs with very large central charge $c$, due to the semi-classical nature of black hole thermodynamics but mostly due to immense analytical simplification it provides. Secondly we will assume a sparse operator spectrum at low scaling dimensions, this allows one to find a kinematical regime where the dominant contribution to correlation function is due to the vacuum Virasoro block, as far as we know there exist no examples of theories with gravitational duals that violate this assumption. 

Similarly we will want to examine the late time behaviour of a set of states that is as broad as possible. One area where the large $c$ approximation particularly shines is the calculation of correlation functions of operators with scaling dimensions that are small with respect to the central charge. Unfortunately in order to overcome the Casimir energy of the Lorentzian cylinder $|E_{C}|=c/24$ one needs to consider states $|V\rangle$ with scaling dimensions $H_V$ proportional to $c$ (typically called heavy states). Put explicitly we will want to consider states of the generic form
\begin{equation}
|V\rangle=g(z)\prod_{k} O_{k}(z_{k})|0\rangle,
\label{states}
\end{equation}
with heavy operators $O_{ik}$ and possibly some sufficiently smooth smearing function $g(z)$. The insertion locations $z_{ik}$ are not completely free since we will want to compute correlators involving this state where there exists an OPE channel where the identity block contribution dominates. One way to achieve this is to restrain $z_{ik}$ to a thin annulus whose exterior boundary is given by the unit circle.

The heaviness of the state provides an additional complication that was tackled by the authors of \cite{Fitzpatrick:2015zha},\cite{Fitzpatrick:2015dlt} in what they call the uniformization approach. They applied there strategy to states generated by acting on the origin with a primary operator, in \cite{Banerjee:2018tut} the authors applied this method to a wider class of heavy states, and showed that while the resulting differential equations typically can not be solved analytically some qualitative features can be derived. It was for instance shown that if the heavy state thermalizes at late time that the temperature matches the temperature of the bulk dual black hole in the Chern-Simons formulation of $AdS_3$ gravity. Some of the results of \cite{Banerjee:2018tut} are briefly covered in section 2.1.

It will be shown that given an integrable heavy state increasing the energy of this state in a controlled way will cause it to cross a threshold that will trigger thermalization at late Lorentzian time. A state that thermalizes at late times will be referred to as a hyperbolic state, conversely states that fail to thermalize will be called elliptic states, this terminology makes particular sense in the following discussion but conveniently (though not coincidentally) this convention was already standardized in \cite{Martinec:1998wm}.

We will see that the existence of a single such threshold is fixed by a combination of representation theory of the Virasoro algebra and classic properties of linear ordinary differential equations.

\subsection{probe two-point function}
To probe the behavior of the final state of an out-of-equilibrium state $|V\rangle$ consider the late-time limit of a two-point function of light operators on the state
\begin{equation}
A(z_1,z_2)=\langle V|Q_h(z_1)Q_h(z_2)|V\rangle
\label{probes}
\end{equation}
the coordinates $z$ are on the radial plane. The operator $Q_h$ is light in the sense that $h\ll c$ as such the amount by which they affect the dynamics of $|V\rangle$ surmounts to a $1/c$ correction, for this reason they form appropriate probes to measure physics of the state $|V\rangle$, in order for the the correlator to be radially ordered we restrict the probe coordinates $z_1, z_2$ to the unit circle. Note that while this correlator is radially ordered on the Euclidean radial plane the Lorentzian correlator of interest is out of time order, this has to be taken into account when analytically continuing to the Lorentzian cylinder. The direct interest of this note is the late Lorentzian time limit, in \cite{Banerjee:2018tut} it was shown that after mapping the radial plane to the Euclidean cylinder $z=\text{exp}(\tau+i\phi)$ and analytically continuing the Euclidean time coordinate $\tau$ time-evolving in Lorentzian time corresponds to letting the probe observers undergo counter-clockwise circles on the radial plane. The non-trivial multi-valuedness of the correlator $A(z_1,z_2)$ ensures that the result is not necessarily periodic in time.

The correlator will be simplified in two directions, first we will only look at the leading order in a $1/c$ expansion secondly we will only consider the identity block exchange, i.e. we only consider the exchange of identity operators when performing OPE contractions between the operators of the state \eqref{states} and the probe operators. Unlike in higher dimensional CFT where the vacuum block contribution merely computes the factorized correlator in 2d CFT the vacuum block contribution to a correlation function captures highly non-trivial information between the interaction of the state onto the probes. One of the Virasoro descendants of the identity operator is the stress tensor, the AdS/CFT dictionary states that its dual field is the graviton field, therefore the bulk interpretation is that Virasoro identity block resums all graviton exchanges between the probes and the geometry generated by the heavy state.

It was found through means closely related to the uniformization of punctured Riemann surfaces \cite{Fitzpatrick:2015zha}\cite{Zograf} that the correlator $A(z_1,z_2)$ could be written in the form
\begin{equation}
A(z_1,z_2)=\psi(z_1)^{-2h_Q}\psi(z_2)^{-2h_Q}\left( \int_{z_1}^{z_2}\psi(z)^{-2}dz\right)^{-2h_Q}
\end{equation}
where $\psi(z)$ satisfies the Fuchsian ODE
\begin{equation}
\psi''(z)+\frac{6}{c}T(z)\psi(z)=0,
\label{Fuchs}
\end{equation}
and where the expectation value $T(z)$ is defined through
\begin{equation}
T(z)=\frac{\langle V|T(z)|V\rangle} { \langle V|V\rangle}.
\label{stresstensor}
\end{equation}
The freedom in picking a particular solution $\psi(z)$ of \eqref{Fuchs} the reflects the freedom in conjoining a uniformizing coordinate system with a global conformal transformation without spoiling its uniformization property. As a complex function $T(z)$ cannot be entirely generic, due to the fact that the in- and out-states are identical up to Hermitian conjugation, taking the complex conjugate of \eqref{stresstensor} shows that it is subject to the following Schwarz reflection property
\begin{equation}
T(z) = z^{-4}T(1/z^*)^*,
\label{reflection1}
\end{equation}
this for one implies that $z^2T(z)$ will be real-valued along the unit circle which will be relevant when we consider the associated Hill system in section 3. Schwarz reflection also restrict the monodromy matrix around the unit circle $M_{\psi}$ of a basis of solutions of \eqref{Fuchs}. There has to exist a basis of solutions such that $M_{\psi}$ is contained within $SU(1,1)$ \cite{Hulik:2016ifr}, which restricts the eigenvalues of $M_{\psi}$ in \textit{any} basis to consist of both purely real numbers or both pure phases. Additionally the determinant constraint fixes the eigenvalues to be each others inverse $(\mu ,1/\mu)$ 

The late Lorentzian time limit corresponds to letting both the coordinates $z_1$ and $z_2$ perform a large number of cycles around the unit circle. If the eigenvalues are given by real numbers $(\mu ,1/\mu)$ the probe correlator is proportional to
\begin{equation}
A\left(e^{2\pi i n_1}z_1,e^{2\pi i n_2}z_2\right)\propto \mu^{-2h(n_1-n_2)}  \hspace{5mm} n_1>n_2\gg1,
\label{latecorrelator}
\end{equation}
identifying $n_i$ with a Lorentzian time through $t_i=2\pi n_i$ we find that \eqref{latecorrelator} resembles a two-point function evaluated on a thermal state with a temperature $T_V$ given by
\begin{equation}
\log(|\mu|)=2\pi^2T_V.
\end{equation}
This concludes a condensed version of some of the results in \cite{Banerjee:2018tut}.

\subsection{Temperature as a conformal invariant and Virasoro orbits}
The arguments reviewed above demonstrate that whether a heavy state $|V\rangle$ thermalizes and what temperature it thermalizes to is controlled by the eigenvalues of the monodromy matrix $M_{\psi}$ of the independent solutions of the linear ODE \eqref{Fuchs}. It can be shown that these eigenvalues are independent of conformal frame, consider the single-valued holomorphic mapping $w(z)$, under this transformation the stress tensor $T(z)$ transforms as
\begin{equation}
T(z)\rightarrow T(w)= \left(\frac{dz}{dw}\right)^2T(z(w))+\frac{c}{12}S[z,w]
\label{Tsubstitution}
\end{equation}
it can easily be verified that if $\psi(z)$ solves the ODE \eqref{Fuchs} that after the substitution \eqref{Tsubstitution} the ODE \eqref{Fuchs} will be solved by
\begin{equation}
\phi(w)=\frac{1}{\sqrt{z'(w)}}\psi(z(w)).
\end{equation}
Therefore the solution $\phi(w)$ under the curve $w(e^{i\th})$ inherits the monodromy of the solution $\psi(z)$ under traversing the unit circle\footnote{There is a small subtlety present concerning winding number, one could construct a map $w(z)$ such that a closed curve in $w$-space has to be traversed multiple times before $z(w)$ performs a single $S^1$ circle in $z$-space. In \cite{Balog:1997zz},\cite{Witten} this issue is bypassed by considering only mappings $S^1\rightarrow S^1$ with winding number 0, in this paper the issue will largely be disregarded.}. Hence the matrix $M_{\psi}$ forms a conformal invariant of the stress tensor.  

A set of functions $T(z)$ that are connected to each other through the action \eqref{Tsubstitution} is given by a Virasoro coadjoint orbit, see appendix A for a short review based on \cite{Balog:1997zz},\cite{Witten},\cite{Kirillov}, for an additional review see \cite{Cotler:2018zff}. All disconnected Virasoro coadjoint orbits have been classified, for instance the set of all hyperbolic orbits is designated by $\mathcal{B}_n(b)$. Here the index $n$ indicates the number of zeroes of the independent solutions of \eqref{Fuchs} in the diagonal basis, the argument $b$ uniquely parametrizes a certain representative element of the orbit. While all the various Virasoro coadjoint orbits have been classified (see table \ref{orbitclassification2}) not all of them are physically relevant. If we define the holomorphic contribution to the energy the usual way as the zero-mode of the stress tensor
\begin{equation}
E=L_0=\frac{1}{2\pi i} \oint dz \, z T(z),
\end{equation}
it is known that all orbits contain elements with arbitrarily large energies. This is not a problem but it does complicate the process of sorting $T(z)$ into an orbit as it almost completely disconnects the temperature and the energy as quantities. What is problematic is that some orbits possess an energy that is unbounded from below, as one of the assumptions of unitary conformal field theory is that the vacuum state is the lowest energy state in any conformal frame this leads to a contradiction, hence we can dismiss these orbits as being unphysical. 

\begin{table}
\begin{tabular}{l||l|l|l|l}
Orbit &Stabilizer & $SL(2,\mathbb{R})$ class & Bounded energy & Constant representative\\
\hline
\hline
$\mathcal{B}_0(b)$ & $S^1$ & hyperbolic & yes & yes\\
\hline
$\mathcal{B}_{n>0}(b)$ & $T_{n,\Delta}$ & hyperbolic & no & no\\
\hline
$\mathcal{C}(\nu) \; (0<\nu<1)$ & $S^1$ & elliptic & yes & yes\\
\hline
$\mathcal{C}(\nu)$ otherwise & $S^1$ & elliptic & no & yes\\
\hline
$\mathcal{P}_0^{+}$& $S^1$ & parabolic & yes & yes\\
\hline
$\mathcal{P}_1^{-}$& $\tilde{T}_{1,-}$ & parabolic & yes & no\\
\hline
$\mathcal{P}_{n}^{\pm}$ & $\tilde{T}_{n,\mp}$ & parabolic & no & no\\
\hline
$\mathcal{E}_{1}$ & $PSL(2,\mathbb{R})$ & exceptional & yes & yes\\
\hline
$\mathcal{E}_{n\neq 1}$& $PSL^{(n)}(2,\mathbb{R})$& exceptional & no & no\\

\end{tabular}

\caption{Summarized form of the various features of the Virasoro orbits as classified in \cite{Balog:1997zz} and reviewed in appendix A.}
\label{orbitclassification2}
\end{table}

Each Virasoro orbit that possesses a lower bound on its energy also possesses a primary representative, meaning that for any $T(z)$ contained within that orbit there exists a single-valued transformation that takes it to a stress tensor of the form $T(z)=H/z^2$. This is exactly the form of the stress tensor expectation value of a state created by a primary operator with scaling weight $H$ inserted at the origin, $O_{H}(0)|0\rangle$. They are related to the constant representatives of \cite{Witten}, \cite{Balog:1997zz} through the coordinate transformation $z=e^{i\th}$, in both cases they are the representatives of an orbit whose energy saturates the lower bound. The fact that each physical orbit contains such a primary representatives has some implications with respect to the eigenstate thermalization hypothesis \cite{srednicki1999approach}, it implies that any generic heavy state is sharply dominated by a single conformal family, see \cite{Banerjee:2018tut} for a short discussion on this point.

The set of distinct physical orbits are dominated by the hyperbolic $\mathcal{B}_0(b)$ orbits and the elliptic $\mathcal{C}_{\nu}$ orbits with $0<\nu<1$. Besides these the physical spectrum consists of one parabolic orbit $\mathcal{P}_0^{+}$ and an exceptional orbit $\mathcal{E}_1$ respectively corresponding to the BTZ cross-over point and the vacuum. Each of these orbits possesses a primary representatives of the form
$$
\begin{array}{lll}
\mathcal{E}_1 &&T(z)=0\\\\
\mathcal{C}_{\nu} \; \; (0<\nu<1)&& T(z)=\frac{c}{24}(1-\nu^2)\frac{1}{z^2},\\\\
\mathcal{P}_0^{+} && T(z)=\frac{c}{24}\frac{1}{z^2},\\\\
\mathcal{B}_0(b)&& T(z)=\frac{c}{24}(1+b^2)\frac{1}{z^2}.\\\\
\end{array}
$$
From the AdS/CFT dictionary \textit{mass} $\longleftrightarrow$ \textit{scaling dimension} it can be seen that it is consistent to attribute to a hyperbolic state a bulk interpretation as being conformally related to a BTZ black hole in the center of AdS. Similarly we can think of elliptic orbits as being conformally equivalent to sourcing a conical defect geometry and of the parabolic orbit as the minimal mass BTZ geometry.

As these are the lowest energy representatives they solidify the CFT intuition that $|V\rangle$ can be thought of as being dominated by a single descendant state of Virasoro representation whose highest weight state has a scaling dimension given by the numerical factor in front of the primary representative stress tensor.

One would rightly think that it would be very difficult to couple a generic state $|V\rangle$ to the appropriate primary representative since they are related to each other by the solutions of extremely complex differential equations \cite{Balog:1997zz}. Fortunately there exists a large body of known theorems about ODEs closely related to \eqref{Fuchs}. These theorems will allow us to prove that we can change the primary representative of a state by acting on a state with a primary operator at the origin, $|V\rangle_H=O_H(0)|V\rangle$.  Given a set of stress tensor expectation values
\begin{equation}
\left\{T_H(z)\right\}=\left\{ \frac{\langle V|O_H(\infty)T(z)O_H(0)|V\rangle}{\langle V|O_H(\infty)O_H(0)|V\rangle} \;\; | \;\; \textrm{All primary $O_H$ in the CFT spectrum}\right\},
\end{equation}
we can think of this family as a curve through the space of all possible stress tensor expectation values parametrized by $H$. In section 3 it will be proven that this curve intersects a bounded Virasoro coadjoint orbit only once, by showing that the constant representative of the orbit increases monotonically as a function of $H$. In fact this tells us that given an elliptic state $|V\rangle$ there exists a certain state-specific critical scaling dimension $H_*$ such that $|V\rangle_{H_*}$ falls within the class of hyperbolic states.

The bulk interpretation is that without any matter in the center of AdS an elliptic orbit $T(z)$ would source a geometry that would evolve to a non-thermal geometry (e.g. a conical defect), but if there was a point source of matter in the center of AdS before releasing $|V\rangle$ then the total scattering mass might be sufficient to trigger black hole collapse.

Finding the exact value for the energy neccessary to trigger such a transition is a very difficult problem but it is possible to determine a set of bounds on the transition energy in terms of the Virasoro conserved charges of the stress tensor. In section 4 it will be derived that for any integer $n\neq 0$ the transition value of $L_0$ designated $L_{|V\rangle}$ is bounded from below by
\begin{equation}
L_{|V\rangle} \geq \frac{c}{24}(1-n^2)+|\text{Re}(L_{-n})|
\end{equation}
From this one can easily interpret this instead as an infinite set of neccessary conditions for a stress tensor $T(z)$ to be contained within a hyperbolic orbit
\begin{equation}
L_0>\frac{c}{24}(1-n^2)+|\textrm{Re}(L_{-n})|,
\end{equation}
to emphasize the meaning of these inequalities, \textit{all} of these need to hold in order for a state $|V\rangle$ associated to $T(z)$ to thermalize, if just one of them fails it implies that $T(z)$ is contained within an elliptic orbit and hence $|V\rangle$ fails to thermalize.

\section{From Fuchs to Hill's equation and its implications}
In order to determine the late-time fate of a state $|V\rangle$ one needs to know the monodromy of the solutions to \eqref{Fuchs} around the unit circle. This suggests that for the purpose of determining the monodromy, or (as we shall see) equivalently the orbit of $T(z)$, knowing the solutions on the full complex plane is largely superfluous. Therefore it makes sense to restrict ODE to the unit circle, as a result one obtains a real-valued ODE of Hill class. The class of Hill's equations are of the following general form
\begin{equation}
y''(x)+\left(\lambda+Q(x)\right)y(x)=0,
\label{hilleq}
\end{equation}
where the coefficient function $Q(x)$ is a periodic function satisfying
\begin{equation}
Q(x+\pi)=Q(x), \hspace{5mm} \int_0^{\pi} dx \, Q(x)=0.
\end{equation}
Furthermore it is assumed that $Q(x)$ is a real function on the interval $[0,\pi]$. There exists a very substantial library of known theorems on the qualitative behavior of this large class of ODEs \cite{MW} which will be exploited in the upcoming sections. Turning our attention back to the Fuchs equation
\begin{equation}
\psi''(z)+\frac{6}{c}T(z)\psi(z)=0,
\end{equation}
where now $T(z)$ is complex meromorphic function subject to the reflection constraint
\begin{equation}
T(z)=\frac{1}{z^4}T(1/z^*)^*,
\label{reflection}
\end{equation}
as mentioned before this directly implies that the product $z^2T(z)|_{|z|=1}$ is real-valued. To rewrite this ODE into Hill form first map the unit circle to the interval $[0,\pi]$ by means of $z=e^{2ix}$, in which case the ODE transforms to
\begin{equation}
\psi''(x)-2i\psi'(x)-\frac{24}{c}e^{4ix}T(x)\psi(x)=0.
\end{equation}
This can (almost) be brought into the canonical form above by means of the transformation $y(x)=e^{-ix}\psi(x)$
\begin{equation}
y''(x)+\left(1-\frac{24}{c}e^{4ix}T(x)\right)y(x)=0.
\end{equation}
In order to fully connect to the canonical form \eqref{hilleq} the zero-mode needs to be subtracted out of the coefficient function
\begin{equation}
\lambda=\frac{1}{\pi}\int_0^{\pi} dx\, 1-\frac{24}{c}e^{4ix}T(x) = 1 - \frac{12}{\pi i c} \oint dz\, zT(z) = 1 - \frac{24}{c}L_0.
\end{equation}
Hence we obtain the Hill system \eqref{hilleq} whose elements are described in table \ref{Hillparameters}.

\begin{table}[h]
\centering
\begin{tabular}{|l|}
\hline
$y(x)=e^{-ix}\psi(x)$ \\
\hline
$Q(x)=\frac{24}{c}\left(L_0-e^{4ix}T(x)\right)$\\
\hline
$\lambda=1 - \frac{24}{c}L_0$\\
\hline
$z=e^{2ix}$\\
\hline
\end{tabular}
\caption{The constituents of the Hill system in terms of CFT data.}
\label{Hillparameters}
\end{table}

There are a few interesting things here to note. First that this is exclusively a well-defined Hill system if the coefficient function is guaranteed to be real for all real values of $x$. This in turn is guaranteed by the reflection symmetry of the stress-tensor expectation value $T(z)$ \eqref{reflection} which implied $z^2T(z)|_{|z|=1}$ is real-valued.

Secondly the number $\lambda$ which plays a very important role in various stability theorems of the Hill equation, has a clear physical interpretation, being for one related to the holomorphic Euclidean energy $L_0$. Additionally if the total energy of the state is converted into a black hole (i.e. no boundary gravitons) $\lambda$ satisfies $\sqrt{|\lambda|}\propto T_{BH}$ \cite{Banerjee:2018tut}\cite{Fitzpatrick:2015zha}. In the next section we will further expand on the CFT role of $\lambda$.


\subsection{The off-set parameter as a CFT quantity}
In a large amount of the theorems we will quote a deciding role will be played by the off-set parameter $\l$. Within the CFT context the parameter $\l$ is not so esoteric as will be shown. Consider a state $|V\rangle$ and its stress tensor expectation value $T(z)$
\begin{equation}
T(z)=\frac{\langle V|T(z)|V\rangle}{\langle V|V\rangle}.
\end{equation}
From the state $|V\rangle$ we can construct a family of states by acting with primary operators
\begin{equation}
|V\rangle_{H}=O_H(0)|V\rangle
\end{equation}
whose holomorphic stress tensor expectation value is given by 
\begin{equation}
T_H(z)\equiv \frac{\langle V|O_H(\infty)T(z)O_H(0)|V\rangle}{\langle V|V\rangle}.
\end{equation}
Both the Fuchs and Hill equation take as input the expectation value of the stress-tensor on the Euclidean radial plane, hence in this specific context no analytic continuation to Lorentzian time is required, as a result the apparent breakdown of radial ordering is of no concern. While no general proof is known it is generally believed that in the large $c$ domain correlators take the general form $e^{-\frac{c}{6}f(z_i)}$ \cite{Hartman:2013mia,Fitzpatrick:2014vua,Zamolodchikov,Zamolodchikov:1995aa}, this combined with the holomorphic Ward identities fixes $T_H(z)$ to the form
\begin{equation}
T_H(z)=\frac{H}{z^2}+\frac{c_V}{z}+T(z),
\end{equation}
demanding that both $T(z)$ and $T_H(z)$ satisfy the reflection symmetry \eqref{reflection1} sets the coefficient $c_V=0$, hence
\begin{equation}
T_H(z)=\frac{H}{z^2}+T(z).
\end{equation}
This demonstrates that acting on a state with a primary operator has the effect of shifting the $L_0$ mode of the stress tensor without affecting the other modes. From the table \ref{Hillparameters} we can as a result read off that acting with a primary operator on the state has the effect of changing the relevant Hill equation to one with the exact same potential $Q(x)$ but with a lower off-set parameter $\l$.

Acting with a primary operator on the state has a clear bulk interpretation, depending on the scaling dimension $H$ it corresponds to creating in the asymptotic past either a conical defect or a black hole in the center of $AdS_3$. In fact one could feasibly construct an experiment in which the value of $\l$ is varied. First prepare an ensemble of systems in various energy eigenstates then at a fixed reference create the collapse state $|V\rangle$ on top of the eigenstate by acting with the operators that create $|V\rangle$ out of the vacuum.

\subsection{Floquet's theorem and its implications}
The main advantage of rewriting the Fuchs equation into Hill form is that it provides access to the general theorems that apply to Hill's equation. Some of these classic results will be quoted without proof in this section, these have all been taken from \cite{MW} which in addition contains all the missing proofs. Some definitions are required, given a Hill system \eqref{hilleq} and a set of normalized solutions $y_1(x)$ and $y_2(x)$ such that
\begin{equation}
y_1(0)=1, \; y_1'(0)=0, \; y_2(0)=0, \; y_2'(0)=1.
\end{equation}
We can define a characteristic equation
\begin{equation}
\rho^2-\left(y_1(\pi)+y_2'(\pi)\right)\rho+1=0,
\label{characteristic}
\end{equation}
whose roots we denote by
\begin{equation}
\rho_1=e^{i\alpha \pi}, \; \rho_2=e^{-i\alpha \pi}.
\end{equation}
after these definitions we can state Floquet's theorem

\begin{theorem}
\textbf{Floquet's theorem} If the roots $\rho_1$ and $\rho_2$ are distinct then Hill's equation has two linearly independent solutions
\begin{equation}
f_1(x)=e^{i\alpha x}p_1(x), \; f_2(x)=e^{-i\alpha x}p_2(x),
\label{floquetfactor}
\end{equation}
where $p_i(x)$ are periodic functions with period $\pi$. If $\rho_1=\rho_2$ then Hill's equation has a periodic solution with either period $\pi$ (when $\rho_1=\rho_2=1$) or period $2\pi$ (when $\rho_1=\rho_2=-1$). Let $p(x)$ denote this periodic solution and let $y(x)$ be another linearly independent solution. Then
\begin{equation}
y(x+\pi)=\rho_1 y(x)+\theta p(x),
\end{equation}
with a constant $\theta$. The case $\theta=0$ corresponds to 
\begin{equation}
y_1(\pi)+y_2(\pi)=\pm 2, \; y_2(\pi)=0, \; y_1'(\pi)=0.
\end{equation}
\end{theorem}
The Floquet basis of solutions is equivalent to the eigenbasis of solutions discussed in section 2, this theorem is essentially a slightly more powerful version of the statement that the monodromy matrix around the unit circle of the solutions of the Fuchs equation is restricted to the set of $SU(1,1)$ matrices. It shares the same conclusion that, depending on wether \eqref{characteristic} has real solutions, the ODE solutions can broadly be divided into a class of solutions that grows/shrinks exponentially and a class that stays bounded at all times. The former will be referred to as unstable solutions and the latter as stable solutions. This brings us to the second theorem, which tells us how the unstable and stable solutions are distributed over parameter space
\begin{theorem}
\textbf{Oscillation theorem} To every differential equation \eqref{hilleq}, there belong two monotonically increasing infinite sequences of real numbers
$$
\lambda_0, \lambda_1, \lambda_2,...
$$
and 
$$
\lambda'_1, \lambda'_2,...
$$
The differential equation \eqref{hilleq} possesses a solution of period $\pi$ if and only if $\l=\l_n$, and a solution of period $2\pi$ if and only if $\l=\l'_n$. The sequences satisfy the inequalities
\begin{equation}
\l_0<\l'_1\leq \l'_2 <\l_1 \leq \l_2 <\l_3'<\l_4'<\l_3<\l_4....
\end{equation}
The solutions of \eqref{hilleq} are stable in the intervals
$$
(\l_0,\l'_1), (\l'_2,\l_1), (\l_2,\l'_3),...
$$
At the endpoints of these intervals the solutions are generally unstable. This is always true for $\l=\l_0$. The solutions are stable for $\l=\l_{2n+1}$ or $\l=\l_{2n+2}$ if and only if $\l_{2n+1}=\l_{2n+2}$, and they are stable for $\l=\l'_{2n+1}$ or $\l=\l'_{2n+2}$ if and only if $\l'_{2n+1}=\l'_{2n+2}$.
\end{theorem}
If we promote a basis of solutions to functions of $x$ and $\l$ as $y_i(x,\l)$ then the characteristic values solve the equations
\begin{eqnarray}
& y_1(\pi,\l_n)+y_2'(\pi,\l_n)=2,\\
& y_1(\pi,\l_n')+y_2'(\pi,\l_n')=-2.
\end{eqnarray}
Due to the fact that $\l=1-24L_0/c$, the oscillation theorem is consistent with the physical intuition that if the energy is taken to infinity while keeping all the other Virasoro conserved charges fixed the state equilibriates to an unstable (black hole) phase with an arbitrarily high temperature. If we now think of the energy as a free parameter in the spirit of primary operator insertions presented in the last section then we can see that lowering the energy causes the solutions to settle down into a stable phase at some point $\l_0$. In principle the oscillation theorem states that instabilities can reoccur if the energy is lowered further but this possiblity will be shown to be excluded.

In order to connect the stability intervals of the oscillation theorem to the Virasoro coadjoint orbits we need one more theorem
\begin{theorem}
Either all nontrivial real solutions of \eqref{hilleq} have only a finite number of zeroes, or all real solutions have infinitely many zeroes. Let $\l_0$ be the smallest value of $\l$ for which \eqref{hilleq} has a periodic solution. Then for $\l \leq \l_0$ all nontrivial real solutions have only a finite number of zeroes, but for $\l>\l_0$, every real solution has infinitely many zeroes.
\end{theorem}
This theorem hints at the fact that the zeroth instability interval is special since it is the only one where solutions possess a finite number of zeroes.

\subsection{Equivalence to Virasoro coadjoint orbits}
At this point two ways of classifying stress tensors have been discussed, one can either sort $T(z)$ into one of the Virasoro coadjoint orbits or one can try to find out in what (in)stability interval its associated Hill system falls. In this section it will be shown that both approaches are equivalent. 

The basis of solutions to the Fuchs equation that diagonalizes the monodromy matrix around the unit circle corresponds to the Floquet basis of solutions of the Hill system. From this we can determine that the hyperbolic class of Fuchs equations, those with eigenvalues $|\mu_i|\neq 1$ correspond to unstable solutions of the Hill system. The set of stress tensors $T(z)$ that lead to hyperbolic Fuchs equations is given by the set of hyperbolic Virasoro coadjoint orbits $\mathcal{B}_n(b)$. The integer $n$ counts the number of zeroes of the solutions of the Fuchs equation on the unit circle \cite{Balog:1997zz} in the diagonal basis, hence for the Hill system the number $n$ corresponds to the number of zeroes on the interval $(0,\pi)$ of the Floquet basis.

Theorem 3 states that the only instability interval whose solutions can have a finite number of zeroes is the zeroth instability interval $\l\in(-\infty, \l_0)$. The solutions associated to the $\mathcal{B}_0(b)$ orbit stress tensors possess no zeroes in the Floquet basis, therefore they have to be contained within the zeroth instability interval. Conversely, all orbits $\mathcal{B}_{n>0}(b)$ lead to Hill solutions that possess at least one zero on the interval $(0,\pi)$. because of the real-valuedness of the Floquet factors $\rho_i$ this zero has to be attributed to the periodic parts $p_i(x)$, hence on the entire real line these solutions have an infinite number of zeroes and they have to be contained within the higher instability intervals. 

On physical grounds all orbits $\mathcal{B}_{n>0}(b)$ can be dismissed as they correspond to states which are conformally related to states with an energy lower than the vacuum. \textit{This implies that all instability intervals other than the zeroth interval can be dismissed as being unphysical.}

From this we can easily establish that all stability intervals other than the first first are similarly unphysical. Simply take such a hypothetical state $|V\rangle$ such that its associated Hill system falls in one of the higher stability intervals. Then we can construct a new state $|V\rangle_H=O_H(x)|V\rangle$ by acting with a primary operator. As derived in section 3.1 this will only affect the Hill system by lowering the value of $\l$. Therefore if we are in a higher stability interval we can pick a scaling dimension $H$ such that $|V\rangle_H$ is contained within a higher instability interval. As a result we conclude that $O_H(0)|V\rangle$ is an unphysical state and cannot be contained within the Hilbert of our CFT which implies that $|V\rangle$ cannot be contained within the physical Hilbert space either.

To complete the equivalence one needs to show the converse statement, that every point on zeroth instability interval and first stability interval respectively correspond to a unique $\mathcal{B}_0(b)$ orbit or $\mathcal{C}_{0<\nu<1}$ orbit. The proofs for these statements are quite a bit more technical and as a result have been relegated to appendix B. The underlying idea is straightforward though, there exists a simple bijective mapping of the Floquet factor $\alpha$ in $e^{i\alpha x}$ and the constant representative of the orbit, in appendix B it is shown that on the zeroth instability interval and the first stability interval the Floquet factor is a monotonic function of $\lambda$, which interpolates every single possible Floquet factor. As a result every single constant representative is hit exactly once on the interval $\l\in(-\infty,\l_1')$.

In short we conclude that every physical bounded Virasoro coadjoint orbit has to correspond to a point on the zeroth instability interval or the first instability interval, and conversely that every point on these two (in)stability intevals in a one-to-one manner exhausts the full set of $\mathcal{B}_0(b)$, $\mathcal{C}_{0<\nu<1}$ and the $\mathcal{P}_0^+$ orbits. This completes the proof for the proposed equivalence between physical orbits and the first two (in)stability intervals. As a direct corollary this provides a CFT version of the fact that there can only be one non-thermal to thermal gravitational phase transition in the bulk.

\section{Putting upper bounds on the transition energy}
Virasoro representation theory restricts the set of physically relevant characteristic values to $\l_0$ (other than $\l_1'$ which will not be discussed). When $\l_0$ is converted into an energy scale $L_{|V\rangle}$ through $\l_0=1-\frac{24}{c} L_{|V\rangle}$ it corresponds to the energy at which the family of states 
\begin{equation}
\left\{|V\rangle_{H}\right\} = \left\{  O_H(0)|V\rangle \; |\,\textrm{All primary $O_H$ in the CFT spectrum}\right\},
\end{equation} 
crosses over to a state that eventually thermalizes at late time. Alternatively one can think of the energy scale $L_{|V\rangle}$ as a diagnostic tool that lets one easily determine whether a state $|V\rangle$ will thermalize or remain in an integrable phase. Unfortunately determining $\l_0$ analytically is an extremely difficult task, either given by the solution to a path-ordered integral \cite{Banerjee:2018tut}, or by the solution of an infinite determinant problem \cite{MW}. In this section we will instead derive a set of upper bounds for the transition energy $L_{|V\rangle}$ in terms of the Virasoro conserved charges. For this purpose consider the following eigenvalue problem which can be constructed out of Hill's equation
\begin{equation}
\left(\frac{d^2}{dx^2} + \l + Q(x)\right)y_n(x)=\sigma_n y_n(x), \hspace{4mm} y_n(x+\pi)=y_n(x),
\end{equation}
from the oscillation theorem we know that if $y(x)$ is constrained to have periodic boundary conditions $y(x+\pi)=y(x)$ that the eigenvalues $\sigma_n$ have to be related to the characteristic values through $\sigma_n=\l-\l_n$. If we are given a stress tensor in the thermal phase we know from the arguments above that $\l<\l_0$. Hence in the thermal phase all eigenvalues are negative and 
\begin{equation}
\textbf{H}(x)\equiv\frac{d^2}{dx^2}+Q(x)+\lambda
\end{equation}
is a negative definite operator. Assuming that the set of periodic solutions $y_n(x)$ forms a complete set on the space of periodic functions we can derive the following inequality
\begin{equation}
\int_0^{\pi} dx \, w(x)\textbf{H}(x)w(x) \leq 0,
\label{HamiltonianBound}
\end{equation}
which should hold for any real-valued periodic function $w(x)$. We can think of $(-)\textbf{H}(x)$ as a Hamiltonian of some quantum system in which case estimating $\l_0$ corresponds to estimating the ground state energy. Note that the inequality \eqref{HamiltonianBound} suggests that one possible strategy to sort $T(z)$ into an orbit would be to check whether there exists a periodic function $w(x)$ such that the inequality \eqref{HamiltonianBound} fails to hold. Mind though that if $T(z)$ is contained within the elliptic phase that $\textbf{H}(x)$ will only have one positive eigenvalue, hence trial-and-error is extremely unlikely to produce a counter-example.

By exploiting the fact that the spectrum of $\textbf{H}(x)$ is bounded from above a more general set of bounds can be derived. This can be done by applying the variational method to estimate the ground state energy of a Hamiltonian. Since $\sigma_0$ is the largest eigenvalue of $\textbf{H}(x)$ the following inequality has to hold for any mormalized periodic function $w(x)$
\begin{equation}
\int_0^{\pi} dx \, w(x)\textbf{H}(x)w(x)\leq \sigma_0.
\end{equation}
Take the following normalized set of test functions $w_n(x)=\sqrt{\frac{2}{\pi}}\sin(nx)$.
\begin{equation}
\int_0^{\pi} dx \, w_n(x)\textbf{H}(x)w_n(x)=\frac{1}{2}(1-n^2)-\frac{12}{c}L_0+\frac{6}{c}\left(L_n+L_{-n}\right),
\end{equation}
To obtain this expression the identities $Q+\lambda=1-\frac{24}{c}e^{4ix}T(x)$ and $L_n=\frac{1}{2\pi i}\oint dz \, z^{n+1}T(z)$ have been used. Since the largest eigenvalue of $\textbf{H}(x)$ is given by $\sigma_0$ we can establish the following general upper bound 
\begin{equation}
1-n^2-\frac{24}{c}L_0+\frac{12}{c}\left(L_n+L_{-n}\right)\leq \sigma_0,
\end{equation}
reinstating $\lambda_0$ through $\sigma_0=\l-\l_0=1-\frac{24}{c}L_0-\lambda_0$ and similarly converting $\l_0$ into the transition energy scale $\l_0=1-\frac{24}{c}L_{|V\rangle}$ yields
\begin{equation}
\frac{c}{24}(1-n^2)+\frac{1}{2}\left(L_n-L_{-n}\right)\leq L_{|V\rangle}.
\end{equation}
This provides us with an infinite set of lower bounds on the black hole transition energy. Therefore if $L_0$ is smaller than the left hand-side for any value of $n$ we know for a fact that the energy of the state $|V\rangle$ is below the transition energy and we can conclude that the stress tensor has to be contained within an elliptic orbit, or to phrase it quantitatively, if there exists any $n$ such that
\begin{equation}
L_0 < \frac{c}{24}(1-n^2)+\frac{1}{2}\left(L_n+L_{-n}\right)
\label{sincriterion}
\end{equation}
then $T(z)$ has to be some element of one of the $C_{\nu}$ elliptic orbits.


By considering the different set of test functions $w_n(x)=\sqrt{\frac{2}{\pi}}\cos(nx)$ a similar set of lower bounds can be obtained\footnote{One might be concerned that this allows test functions that posses test functions with period $2\pi$ instead of $\pi$, the oscillation theorem states that Hill's equation can only possess non-trivial solutions with period $2\pi$ at the characteristic values $\l'_n$. Since $\l_0<\l'_n$ for all $n$ this does not affect the conclusion.}
\begin{equation}
\frac{c}{24}(1-n^2)-\frac{1}{2}\left(L_n-L_{-n}\right)\leq L_{|V\rangle}.
\label{coscriterion}
\end{equation}
The interpretation of an orbit as a classical Virasoro representation suggests that the $L_n$ charges are due to lowering operators and as such the dependence on these numbers is in a sense redundant, the reflection condition \eqref{reflection} imposes that $L_{n}=L_{-n}^*$ this allows us to eliminate the lowering operator charges. Furthermore the two sets of lower bounds \eqref{sincriterion} and \eqref{coscriterion} can be combined to give
\begin{equation}
\frac{c}{24}(1-n^2)+|\text{Re}(L_{-n})|\leq L_{|V\rangle}.
\end{equation}


This provides the following general necessary conditions for whether a state will thermalize
\begin{equation}
L_0>\frac{c}{24}(1-n^2)+|\text{Re}(L_{-n})|,
\label{necessary}
\end{equation}
these are necessary in the sense that the state $|V\rangle$ will exclusively thermalize if all of these condtions hold. The major downside is of course that these are only necessary conditions not sufficient conditions. The CFT interpretation is that $|V\rangle$ is generically not a primary state but a descendent. Only the contribution to the energy from the primary contributes to the constant representative. These criteria tell us whether, after filtering out the energy contribution from a particular Virasoro raising mode, the energy is still sufficient to overcome the holomorphic Casimir energy equal to $c/24$. Alternatively one can think of the inequalities \eqref{necessary} as bounding an equation of state in the thermalizing phase if the Virasoro conserved charges are interpreted as thermodynamic equation of state variables.


\section{Discussion}
The universality of black hole formation and black hole thermodynamics of gravity theories on asymptotic $AdS_3$ suggests that the dual description of thermalization should equally be fixed by the universal features of 2d conformal field theory. This paper set out to demonstrate that a similar transition to a thermal state is fixed by a combination of Virasoro representation theory, properties of linear differential equations and the universal vacuum block contribution to correlators. As a consequence a set of upper bounds were derived that firstly let one, in some cases, determine if a given heavy state is elliptic (i.e. non-thermal) and secondly provides lower bounds on the deficient energy required to trigger thermalization.

To probe thermalization it is typical to compute correlators of a set of simple operators and see if, to leading order in a statistical mechanics large system limit, they match thermal expectation values. This paper specifically considers two-point functions of probe operators, these operators have small scaling dimensions compared to the system. Therefore they can be interpreted as corresponding to low-energy excitations in radial quantization. In the case considered the large central charge expansion plays the role of the large system limit. In principal the number of probe operators can be scaled up relatively straightforwardly as long as the resulting product operator is parametrically light compared to the central charge. Physically this correponds to the fact that a sufficiently complicated quantum measurement can resolve the microstate of the ensemble which in the process destroys the statistical mechanics interpretation. Quantitatively the Virasoro Ward identity states that one would have to sum up all light contributions to the stress tensor expecation value if there are order $O(c)$ of these $O\left(c^0\right)$ contributions the large $c$ expansion will break down.

Intuitively black holes formed by dynamical collapse would be an ideal arena to study the information paradox. In fact since the leading order large central charge calculation reproduces the correlator decay discussed in Maldacena's formulation of the information paradox \cite{Maldacena:2001kr}, it is tempting to compute perturbative corrections. Especially since the authors of \cite{Fitzpatrick:2015dlt} showed that these corrections can be computed systematically. Unfortunately almost exactly the same authors demonstrated that these perturbative corrections can not resolve the information paradox \cite{Fitzpatrick:2016ive}, see also \cite{Kraus:2018zrn} for a recent discussion. Perturbative $1/c$ corrections are not entirely devoid of practical meaning though, and are in fact very efficient for the purpose of calculating gravitational loop corrections. The first $1/c$ corrections of the probe part contributes the Eikonal resummation of ladder diagrams of graviton exchanges between the probe fields and higher order corrections resum loop corrections on top of that.

\section*{Acknowledgements}
The author of this paper would especially like to thank K. Papadodimas for the useful discussions and his general patience. Furthermore for other discussions the author would like to thank S. Banerjee and J. Sonner. This work has been supported by the Royal Netherlands Academy of Arts and Sciences (KNAW).



\appendix

\section{Appendix: Virasoro coadjoint orbits}
Due to their prevalence in the main body of the text it would be useful to review some of the basic facts of Virasoro coadjoint orbits. First we will review that the Lie derivatives of orientation preserving diffeomorphisms of the circle (Diff($S^1$)) form an algebra that in the fourier basis gives the Witt algebra. As a result the Virasoro algebra has a geometrical meaning as the central extension of the Lie algebra of Diff($S^1$). From this we will be able to read off that the CFT stress-tensor has a geometrical interpretation in terms of a quadratic differential field on a manifold whose tangent bundle is given by the (extended) family of infinitessimal elements of Diff($S^1$). \cite{Balog:1997zz,Lazutkin,Garbarz:2014kaa}

Since $S^1$ is 1-dimensional the algebra constructed out of their Lie derivatives is particularly simple. Given two vector fields on $S^1$, $V=f(\theta)\partial_{\theta}$ and $W=g(\theta)\partial_{\theta}$.
\begin{equation}
[W,V]=\mathcal{L}_{W}V=\left(f'g-g'f\right)\partial_{\theta}.
\end{equation}
If we decompose the functions $f(\theta)$, $g(\theta)$ in Fourier modes 
\begin{equation}
f(\theta)=\sum_n i L_n e^{i\theta (n+1)},
\end{equation}
and invert it we obtain the Witt algebra for the modes $L_n$
\begin{equation}
[L_n,L_m]=(n-m)L_{n+m}.
\end{equation}
As is well-known, the Virasoro algebra is the central extension of the Witt algebra. As such we have to motivate a particular way to extend this algebra. The elements are easy, the basic elements of Witt algebra were the tangent vector fields on the circle parametrized by functions $f(\theta)$. To extend the elements we simply append to each function $f(\theta)$ a number to create a doublet $(f(\theta),t)$, these will be the elements of the extended algebra. The complication is to find a new bracket on this extended algebra, i.e given two doublets $(f(\theta),t_1)$, $g(\theta),t_2)$ what is
$$
[(f(\theta),t_1),(g(\theta),t_2)]=\; ?
$$
The standard CFT approach is to define the bracket through the 2d conformal Ward identity
\begin{equation}
\delta_{\ep}\langle X \rangle = -\frac{1}{2\pi i}\oint dz \, \ep(z)\langle T(z) X\rangle.
\end{equation}
From this Ward identity we can associate the following charge operator to the infinitessimal conformal transformation $\ep(z)$
\begin{equation}
Q_{\ep}[T]=\frac{1}{2\pi} \oint dz \, \ep(z)T(z),
\end{equation}
we associate the following Lie bracket to these conserved charges
\begin{equation}
[Q_{\ep_1},Q_{\ep_2}][T]\equiv -Q_{\ep_2}[\delta_{\ep_1}T].
\label{chargebracket}
\end{equation}
We know the effect of the infinitessimal transformation on the stress tensor
\begin{equation}
\partial_{\ep}T(z)=-\frac{c}{12}\partial_z^3\ep(z)-2T(z)\partial_z \ep(z) - \ep(z)\partial_z T(z),
\end{equation}
hence we can work out the right-hand side of \eqref{chargebracket}
\begin{multline}
[Q_{\ep_1},Q_{\ep_2}][T]= \frac{1}{2\pi} \oint dz \, \ep_2(z)\left(-2T(z)\partial_z\ep_1(z)-\ep_1(z)\partial_zT(z)-\frac{c}{12}\partial_z^3\ep_1(z)\right)\\\\=\frac{1}{2\pi}\oint dz \, \left(\ep_1(z)\partial_z\ep_2(z)-\ep_2(z)\partial_z\ep_1(z)\right) T(z) - \frac{c}{24\pi}\oint dz \, \ep_2(z)\left(\partial_z^3 \ep_1(z)\right)\\\\=Q_{[\ep_1,\ep_2]}[T(z)] - \frac{c}{24\pi}\oint dz \, \ep_2(z)\left(\partial_z^3 \ep_1(z)\right).
\end{multline}
Note that the boundary term coming from the integration by parts vanishes due to the assumed single-valuedness of $T(z)$ and $\ep_i(z)$. This candidate Lie bracket does not look anti-symmetric yet, this can be fixed by performing two integrations by parts on the anomalous term
\begin{multline}
\oint dz \,  \ep_2(z) \partial_z^3\ep_1(z) = -\oint dz\, \partial_z \ep_2(z) \partial_z^2 \ep_1(z) \\\\ =-\oint dz \, \frac{1}{2}\left( \partial_z \ep_2(z) \partial_z^2 \ep_1(z) + \partial_z \ep_2(z) \partial_z^2 \ep_1(z)\right)\\\\=\oint dz \, \frac{1}{2}\left(\partial_z^2\ep_2(z)\partial_z \ep_1(z)-\partial_z^2\ep_1(z)\partial_z \ep_2(z)\right).
\end{multline}
This commutator algebra on the conformal charges of 2d CFT motivates the following bracket on the extended Witt algebra by transforming back to our coordinate system through $z=e^{i\theta}$. After some algebra we find
\begin{equation}
[(f(\theta),t_1),(g(\theta),t_2)]=\left([f(\theta),g(\theta)]\,,\; \frac{1}{48\pi}\int_0^{2\pi} d\theta \, e^{-2i\theta}\left(f'(\theta)g''(\theta)-g'(\theta)f''(\theta)\right)\right).
\label{vectorbracket}
\end{equation}
Write the space $(f(\theta)\partial_{\theta},a)$ in terms of the following basis
\begin{equation}
L_n= \left(ie^{i(n+1)\theta}\partial_{\theta},0\right), \; Z=(0,i/2).
\end{equation}
It is then easy to check that we find the usual Virasoro algebra (in a slightly unconventional form)
\begin{eqnarray}
& [L_n,L_m]=(n-m)L_{n+m}+\frac{Z}{12}(n+1)^3\delta_{n+m,0},\\
& [L_n,Z]=0.
\end{eqnarray}
The additional generator $Z$ commutes with all other generators, hence this is a proper central extension of the Witt algebra.

\subsection{Coadjoint action}
In the last subsection it was reviewed how the Witt algebra can be viewed as the Lie bracket of vector field on $S^1$ and the Virasoro algebra as a central extensions of this algebra. To define the adjoint and coadjoint action on these vector fields we need to define an inner product. The vector fields $f(\theta)\partial_{\theta}$ and quadratic differential fields $b(\th)d\th^2$ possess a natural inner product
\begin{equation}
\langle b(\th),f(\th)\rangle=\int_0^{2\pi} d\theta \, f(\th)b(\th),
\end{equation}
by simple Cartesian extension of this scalar product we can define an inner product on the extended space
\begin{equation}
\langle (b(\th), a), \, (f(\th),t)\rangle = \int_0^{2\pi} d\theta \, f(\th)b(\th) \, + \, at.
\end{equation}
The adjoint representation is defined through the action of the vector fields onto themselves i.e.
\begin{equation}
ad_u(v)\equiv [u,v]
\end{equation}
from this definition we can construct the adjoint action through the bracket \eqref{vectorbracket}. Inserting this into the inner product tells us the effect of the adjoint action
\begin{multline}
\langle (b,a), \, ad_{(g,x)}(f,y)\rangle = \langle (b,a), \, \left( f'g-g'f, \, \frac{1}{48\pi}\int_0^{2\pi}d\theta \, e^{-2i\theta}(f'g''-g'f'')\right)\rangle \\\\= \int_0^{2\pi} d\th \, b(f'g-g'f) + \frac{a}{48\pi}\int_0^{2\pi}d\th \, e^{-2i\th}(f'g''-g'f'').
\end{multline}
Given the adjoint action we define the coadjoint action as the unique operator on the dual space such that its effect on inner products is the same as that of the adjoint action\footnote{Up to a sign that is, this sign convention ensure that the combination $\langle ad^{*}_{(g,x)}(b,a), \, (f,y)\rangle + \langle (b,a),\, ad_{(g,x)}(f,y)\rangle$ vanishes.}, specifically
\begin{equation}
\langle ad^{*}_{(g,x)}(b,a), \, (f,y)\rangle \equiv -\langle (b,a),\, ad_{(g,x)}(f,y)\rangle.
\end{equation}
Since the right hand-side is known we can compute what the effect of the coadjoint action is on a quadratic differential pair $(b,a)$
\begin{equation}
\langle ad^*_{(g,x)}(b,a),\, (f,y)\rangle \equiv -\langle (b,a),\, ad_{(g,x)}(f,y)\rangle =-\int_0^{2\pi} d\th \, bf'g-bg'f + \frac{a}{48\pi} e^{-2i\th}(f'g''-g'f''),
\end{equation}
after performing some integrations by parts we obtain
\begin{equation}
\langle ad^*_{(g,x)}(b,a),\, (f,y)\rangle=\int_0^{2\pi} d\theta \, \left(-2bg'-b'g -\frac{a}{48\pi}e^{-2i\theta}\left( 2g'''-6ig''-4g'\right)\right)f,
\label{coadjointaction}
\end{equation}
from which we can read off the effect of the coadjoint action
\begin{equation}
ad^*_{(g,x)}(b,a)=\left(2bg'+b'g +\frac{a}{48\pi}e^{-2i\theta}\left( 2g'''-6ig''-4g'\right),\, 0 \right).
\end{equation}
This expression looks unfamiliar but if we apply the change of variables $z=e^{i\th}$ on the right hand-side of \eqref{coadjointaction} we obtain
\begin{equation}
\langle ad^*_{(g,x)}(b,a),\, (f,y)\rangle=\oint dz \left(2bg'+b'g -\frac{a}{24\pi}g'''\right)f
\label{coadjointactionz}
\end{equation}
hence in $z$-coordinates the coadjoint action takes the form of the variation of the 2d CFT stress-tensor under infinitessimal conformal transformations generated by $g$. This justifies the description maintained throughout the body of the text, that the sets of all stress-energy tensors that are connected to each other through single-valued conformal transformations is isomorphic to the coadjoint orbits generated from a quadratic differential field on the extended space by acting with the coadjoint action.

\subsection{Classifying orbits}
In the previous subsection it was shown that due to the transformation rule of the stress tensor it naturally lives inside a Virasoro coadjoint orbit, i.e. the space of all quadratic differentials that are continuously connected to some reference quadratic differential field. These coadjoint orbits possess a natural manifold structure \cite{Kirillov}, where we can think of the coadjoint action parametrized by a pair $(f,x)$ as (almost) the tangent space at the reference point $(b,a)$. This is not entirely accurate, there are possibly also pairs $(h,y)$ that leave the reference point $(b,a)$ invariant, these have to be modded out. Define the central extension of the algebra of vector fields on $S^1$ as $\overline{diff(S^1)}$ and $G_b$ as the the algebra spanned by the vector fields $(h,y)$ such that $ad^{*}_{(h,y)}(b,a)=0$ specifically
\begin{equation}
G_b=\left\{ h\in \overline{diff(S^1)}\, | \; ad^{*}_h(b,a)=0\right\}.
\end{equation}
In that case the tangent space of the coadjoint orbit at the reference point $(b,a)$ is given by $\overline{diff(S^1)}/H_b$. This tangent space can be integrated to a full manifold designated by $\overline{Diff(S^1)}/H_b$, these will be identified as the coadjoint orbits. As a result the coadjoint orbits are classified by their reference point and their stabilizer algebra. Turning back to the stabilizer algebra, since the coadjoint action is given by
\begin{equation}
ad^*_{(f,x)}(b,a)=\left(2bf'+b'f +\frac{a}{48\pi}e^{-2i\theta}\left( 2f'''-6if''-4f'\right),\, 0 \right).
\end{equation}
we know that given a fixed point $(b,a)$ the defining equation for the stabilizer vectors $h$ is
\begin{equation}
2bh'+b'h +\frac{a}{48\pi}e^{-2i\theta}\left( 2h'''-6ih''-4h'\right)=0,
\end{equation}
or after a change of variables $z=\textrm{exp}(i\th)$
\begin{equation}
2bh'+b'h -\frac{a}{24\pi}h'''=0.
\label{stabilizereq}
\end{equation}
As a first step towards the well-studied \cite{Balog:1997zz} connection between Virasoro coadjoint orbits and the differential equations classified as Hill's equation we note that given a basis of solutions $\psi_1(z)$ and $\psi_2(z)$ to the linear ODE
\begin{equation}
\psi''(z)-b(z)\psi(z)=0,
\end{equation}
the independent solutions of \eqref{stabilizereq} will be given by
\begin{equation}
\psi_1(z)^2, \hspace{3mm} \psi_2(z)^2, \hspace{3mm} \psi_1(z)\psi_2(z).
\label{stabilizersolutions}
\end{equation}
There is one additional complication, the condition \eqref{stabilizereq} is a local condition, there is no guarantee that the local solutions \eqref{stabilizersolutions} respect the periodicity condition of the circle. In a particularly impressive proof in \cite{Witten} it is shown that the number of global stabilizing vector fields of a given reference point is always either 1 or 3.

The classes of solutions to \eqref{stabilizereq} have been classified \cite{Witten}, the resulting (either 1 or 3 dimensional) stabilizer groups that can be uplifted from the solutions are typically listed as \cite{Balog:1997zz}
\begin{equation}
S^1, \hspace{3mm}, PSL^{(n)}(2,\mathbb{R}), \hspace{3mm} T_{n,\Delta}, \hspace{3mm} \tilde{T}_{n,\pm}.
\end{equation}
Respectively, $S^1$ forms the set of rigid rotations of the circle, $PSL^{(n)}(2,\mathbb{R})$ is the group generated from the subalgebra of $\overline{diff(S^1)}$ spanned by $(L_0,L_{-n},L_{n})$. $T_{n,\Delta}$ and $\tilde{T}_{n,\pm}$ are one-dimensional groups whose Lie algebras consist of vector fields on $S^1$ with respectively $2n$ simple zeros or $n$ double zeros, since the number of zeros is an orbit invariant The list of these group manifolds that form the various Virasoro coadjoint orbits has been summarized in table \ref{orbittable} \cite{Garbarz:2014kaa}.
\begin{table}
\begin{tabular}{c||c}
Orbit & Stabilizer algebra generators\\
\hline
\hline
$\overline{Diff(S^1)}/S^1$ & $l_0$\\
\hline
$\overline{Diff(S^1)}/PSL^{(n)(2,\mathbb{R})}$ & $l_0, l_n, l_{-n}$ \\
\hline
$\overline{Diff(S^1)}/T_{n,\Delta}$ & $f(\th)$, where $f$ is a vector field with $2n$ simple roots\\
\hline
$\overline{Diff(S^1)}/\bar{T}_{n,\pm}$ & $f(\th)$, where $f$ is a vector field with $n$ double roots\\
\end{tabular}
\caption{The classification of Virasoro coadjoint orbits purely in terms of group manifolds, i.e. without their reference points} 
\label{orbittable}
\end{table}

\subsection{Hill's equation, Virasoro coadjoint orbits and $SL(2,\mathbb{R})$ conjugacy classes}
So far this concludes the classification as presented in \cite{Witten}, the group manifolds formed by the coadjoint orbits form somewhat intangible structures. In this section we will follow \cite{Balog:1997zz} and link the Virasoro coadjoint orbits to something that is of direct physical interest for our purposes; the conjugacy classes of the monodromy matrices of the solutions $\psi(z)$ of
\begin{equation}
\psi''(z)+\frac{6}{c}T(z)\psi(z)=0,
\label{fuchs}
\end{equation}
around the unit circle. The information of the coadjoint orbit enters the ODE through $T(z)$ which is related to the earlier quadratic differential fields $b(\th)d\th^2$ through $T(z)=b(x(z))$ with $z=e^{i\th}$. In order to directly connect to application in 2d CFT described in section 2 we identify $T(z)$ with the stress tensor expectation value of the CFT used throughout the body of the text, in that vein $c$ designates the central charge of theory. Being a second order linear differential equation, the solution space of \eqref{fuchs} is spanned by a basis of solutions of the form $\pmb{\psi}(z)=(\psi_1(z),\psi_2(z))^{T}$, the monodromy matrix $\pmb{\psi}(z)$ around the unit circle $M_{\psi} \in SL(2,\mathbb{R})$ is given by
\begin{equation}
\pmb{\psi}\left(e^{2\pi i}z\right)=M_{\psi} \pmb{\psi}(z)
\end{equation}

The infinitessimal transformation under a vector field $(f,y)$ was given in \eqref{coadjointactionz} it is known that this expression integrates to the transformation rule
\begin{equation}
T(w)=\left(\frac{dz}{dw}\right)^2T(z)+\frac{c}{12}S[z,w],
\label{Ttransformation}
\end{equation}
where $w(z)$ is an element of $Diff(S^1)$ analytically continued to the complex plane and $S[z,w]$ is the Schwarzian derivative defined as
\begin{equation}
S[z,w]=\frac{z'''(w)}{z'(w)}-\frac{3}{2}\left(\frac{z''(w)}{z'(w)}\right).
\end{equation}
It is straightforward to check that if $\psi(z)$ solves \eqref{fuchs} that under the replacement $T(z)\rightarrow T(w)$ as in \eqref{Ttransformation} that \eqref{fuchs} is solved by the replacement $\psi(z)\rightarrow \phi(w)=\frac{1}{\sqrt{z'(w)}}\psi(z(w))$. Since $w(z)$ is by construction single-valued along the unit circle we can conclude that the monodromy matrix $M_{\psi}$ is invariant under the transformations $w(z)$. Of course in constructing the basis of solutions $\pmb{\psi}(z)$ there exists a freedom in choice of basis, a basis tranformation $\pmb{\psi}_A(z)=A\pmb{\psi}(z)$ has the effect of conjugating the monodromy matrix
\begin{equation}
M_{\psi_A}=A^{-1}M_{\psi}A.
\end{equation}
This establishes an important fact namely that the conjugacy class of the monodromy matrix $M_{\phi}$ is an orbit invariant with respect with the entire Virasoro orbit in which $T(z)$ is contained.

Since every Virasoro orbit maps to a conjugacy class of $SL(2,\mathbb{R})$ and since these conjugacy classes are known, the problem is to link the right orbit to the right conjugacy class. The strategy is to match the respective stabilizer subgroups. Take $\alpha(z)$ to be an element of the stabilizer group of $T(z)$, i.e.
\begin{equation}
T(z)=\left(\frac{dz}{d\alpha}\right)^2T(z)+\frac{c}{12}S[z,\alpha].
\end{equation}
The fact that $\alpha(z)$ leaves the differential equation invariant but changes the solution implies that the new solutions are related to the old ones by linear transformation
\begin{equation}
\frac{1}{\sqrt{\alpha'(z)}}\pmb{\psi}(\alpha(z))=\gamma(\alpha)\pmb{\psi}(z),
\end{equation}
the left hand side leaves $M_\psi$ invariant whereas the right-hand side affects $M_\psi$ through conjugation as a result the matrix $\gamma(\alpha)$ has to satisfy
\begin{equation}
M_\psi=\gamma(\alpha)^{-1}M_{\psi}\gamma(\alpha).
\end{equation}
As a mapping $\gamma(\alpha)$ has the property of mapping of mapping an element of the stabilizer group to a matrix that leaves the matrix $M_\psi$ invariant under conjugation, this set of matrices forms a group $G[M_\psi]$ and $\gamma(\alpha)$ forms a homomorhpism from the stabilizer group to $G[M_\psi]$. This led the authors of \cite{Balog:1997zz} to the following strategy: take a conjugacy class of $SL(2,\mathbb{R})$, pick an element of that class to act as a representative, compute the set of matrices that leave the representative invariant under conjugation. As a final, most computationally challenging, task find a $T(z)$ such that a basis of solutions to \eqref{fuchs} exists such that it's monodromy around the unit circle is given by the representative matrix. As a result one will have found the set of representatives and stabilizer groups hence classifying all Virasoro coadjoint orbits.

The set of eigenvalues of an element of $SL(2,\mathbb{R})$ are invariant under conjugation, the $\textrm{det}(M_{\psi})=1$ constraint restricts the two eigenvalues of $M_{\psi}$ to be each others inverse, Floquet's theorem further restricts the relevant classes to either be purely real or pure phase with cross-over points where both eigenvalues are either 1 or both -1. The orbits with eigenvalues falling in these categories are respectively designated as hyperbolic, elliptic and parabolic or exceptional orbits. By constructing explicit examples of stress tensors $T(z)$ whose associated solutions have a monodromy within these conjugacy classes the authors of \cite{Balog:1997zz} list the following classes of Virasoro orbits
\begin{equation}
\mathcal{B}_n(b), \hspace{3mm} \mathcal{C}(\nu), \hspace{3mm} \mathcal{P}_{n}^{\pm}, \hspace{3mm} \mathcal{E}_n
\end{equation}
these respectively correspond hyperbolic, elliptic, parabolic and exceptional orbits, but they can have different stabilizer algebras corresponding to the affixes $n$ and $\pm$ the real arguments $b$ and $\nu$ parametrize the set of representative stress tensors.

\subsection{Relevant qualitative features of the individual orbits}
As we shall see not all the orbits mentioned in the previous section are of direct physical interest though. To each stress tensor we an associate an energy expectation value through the conserved charge $L_0$
\begin{equation}
L_0=\frac{1}{2\pi i}\oint dz \; zT(z),
\end{equation}
while within each orbit this number is unbounded from above, there are only a few orbits for which this number bounded from below. Since we are considering unitary CFTs the energy expectation value on the entire spectrum of states is bounded from below by the energy expectation value of the vacuum state (in any conformal frame). Therefore one can dismiss any orbit without a lower bound as incapable of being associated to a physical state of the CFT spectrum. The full list of orbits which possess a lower bound on their energy is given by
\begin{equation}
\mathcal{B}_0(b), \hspace{3mm} \mathcal{C}(\nu) \;\; (0<\nu<1), \hspace{3mm} \mathcal{P}_{0}^{+}, \hspace{3mm} \mathcal{E}_1, \hspace{3mm} \mathcal{P}_1^{-}.
\end{equation}
In addition some of these orbits have the special property that they contain constant representatives\footnote{The name is due to fact that in the coordinate $x$ of \cite{Balog:1997zz},\cite{Witten} these representatives are in fact constants, for the 2d CFT purposes of this note it is convenient to immediately express these representative stress tensors in the $z$-coordinate of the radial plane, where constant representative is perhaps a bit of a misnomer. Hence in the main body of the text they are instead referred to as primary representatives.}  these orbits and their representatives are given by
$$
\begin{array}{lll}
\mathcal{E}_1 &&T(z)=0\\\\
\mathcal{C}_{\nu} \; \; (0<\nu<1)&& T(z)=\frac{c}{24}(1-\nu^2)\frac{1}{z^2},\\\\
\mathcal{P}_0^{+} && T(z)=\frac{c}{24}\frac{1}{z^2},\\\\
\mathcal{B}_0(b)&& T(z)=\frac{c}{24}(1+b^2)\frac{1}{z^2}.\\\\
\end{array}
$$
The representative functions $T(z)$ have the form of the stress-tensor expectation values on a state given by conformal primary inserted at the origin acting on the vacuum with a scaling dimension given by the numerical prefactor of $T(z)$. This also serves as a first indicator that something seems to change at the BTZ mass threshold $c/24$. Some of the details of the classification have been summarized in table \ref{orbitclassification}.
\begin{table}
\begin{tabular}{l||l|l|l|l}
Orbit &Stabilizer & $SL(2,\mathbb{R})$ class & Bounded energy & Constant representative\\
\hline
\hline
$\mathcal{B}_0(b)$ & $S^1$ & hyperbolic & yes & yes\\
\hline
$\mathcal{B}_{n>0}(b)$ & $T_{n,\Delta}$ & hyperbolic & no & no\\
\hline
$\mathcal{C}(\nu) \; (0<\nu<1)$ & $S^1$ & elliptic & yes & yes\\
\hline
$\mathcal{C}(\nu)$ otherwise & $S^1$ & elliptic & no & yes\\
\hline
$\mathcal{P}_0^{+}$& $S^1$ & parabolic & yes & yes\\
\hline
$\mathcal{P}_1^{-}$& $\tilde{T}_{1,-}$ & parabolic & yes & no\\
\hline
$\mathcal{P}_{n}^{\pm}$ & $\tilde{T}_{n,\mp}$ & parabolic & no & no\\
\hline
$\mathcal{E}_{1}$ & $PSL(2,\mathbb{R})$ & exceptional & yes & yes\\
\hline
$\mathcal{E}_{n\neq 1}$& $PSL^{(n)}(2,\mathbb{R})$& exceptional & no & no\\

\end{tabular}
\caption{Relevant features of the Virasoro coadjoint orbits as classified in \cite{Balog:1997zz}.}
\label{orbitclassification}
\end{table}

\FloatBarrier

\section{Floquet factor as a monotonic function of $\lambda$}
The number $\lambda$ contained within Hill's equation possesses within our context the interpretation as being related to the total energy of our state on the radial plane through $\l=1-\frac{24}{c}L_0$. It would hence seem sensible that decreasing $\lambda$ would increase the mass of the conformal primary associated to an orbit. In this section we will prove that this is the case, in more practical terms, this establishes that as we decrease $\lambda$ from $\lambda_1'$ to negative infinity we sweep every physically allowed orbit and we reach every orbit exactly once. The scaling dimension $H$ of the primary representative of an orbit $T(z)=H/z^2$ is simply related to the Floquet factor $\alpha$ in \eqref{floquetfactor} through 
\begin{equation}
\alpha=\sqrt{1-24\frac{H}{c}},
\end{equation}
therefore we will argue the equivalent statement that varying $\l$ from $\l_1'$ to negative infinity monotonically interpolates every possible Floquet factor. The proof is somewhat technical, mostly due to the fact that any proof has to naturally exclude the instability intervals that are not the zeroth interval due to the clear fact that the Floquet factor cannot be monotonic within these intervals. The proof for the first stability interval has already been done in the past, in \cite{MW} it is proven that Hill's discriminant is monotonically decreasing within this interval.  

To prove the monotonicity of the Floquet factor in the zeroth instability interval consider the Floquet basis of solutions
\begin{equation}
y_1(x)=e^{-\beta x}p_1(x), \hspace{5mm} y_2(x)=e^{\beta x} p_2(x)  \hspace{8mm} \textrm{with} \hspace{3mm} p_i(x+\pi)=p_i(x),
\label{hill2}
\end{equation}
of Hill's equation
\begin{equation}
y''(x)+\left(Q(x)+\lambda\right)y(x)=0
\end{equation}
an important point to note is that all solutions associated to Hill's equation with $\lambda<\lambda_0$ have only a finite amount of zeroes, the periodic nature of $p_i(x)$ hence implies that $y_i(x)$ has no zeroes. This means that both functions $y_i(x)$ have definite sign for all $x$ and since Hill's equation is linear we can choose this sign to be positive without loss of generality for both Floquet solutions
\begin{equation}
y_i(x)>0 \hspace{5mm} \textrm{for all $x$ as long as $\lambda<\lambda_0$}.
\end{equation}
now consider the following Hill equation shifted by a small $\epsilon$
\begin{equation}
v''(x)+(Q(x)+\lambda-\epsilon)v(x)=0,
\label{hill3}
\end{equation}
this equation will have another Floquet basis
\begin{equation}
v_1(x)=e^{-\gamma x}q_1(x), \hspace{5mm} v_2(x)=e^{\gamma x} q_2(x)  \hspace{8mm} \textrm{with} \hspace{3mm} q_i(x+\pi)=q_i(x),
\end{equation}
The same consideration that applied to $y_i(x)$ apply to $v_i(x)$, i.e. the functions $v_i(x)$ can be taken to be strictly greater than zero for all $x$. The solutions of Hill's equation should continuously flow into each other as we vary $\lambda$. Consider the solution $\tilde{y}_2(x)$ to \eqref{hill3} obtained from the exponentially increasing $y_2(x)$ by continuously shifting $\lambda\rightarrow \lambda -\epsilon$. As a solution it should be expressible as a linear combination of Floquet solutions 
\begin{equation}
\tilde{y}_2(x)= a v_1(x) +b v_2(x) = a e^{-\gamma x} q_1(x) +b e^{\gamma x} q_2(x),
\end{equation}
since $y_2(x)$ is positive everywhere we can choose an $\epsilon$ such that the solution $\tilde{y_2}(x)$ is positive everywhere as well, this implies that both $a,b>0$. Now consider the ratio $\tilde{y}_2(x)/y_2(x)$ in the limit $x\rightarrow \infty$
\begin{equation}
\lim_{x\rightarrow\infty} \frac{\tilde{y}_2(x)}{y_2(x)} = \lim_{x\rightarrow\infty} \frac{a e^{-\gamma x} q_1(x) +b e^{\gamma x} q_2(x)}{e^{\beta x} p_2(x)} = \lim_{x\rightarrow\infty} b e^{(\gamma-\beta)x}\frac{q_2(x)}{p_2(x)}.
\end{equation}
Since $q_2(x)$ is a periodic function and hence bounded and since $p_2(x)$ can vanish nowhere we can conclude that if $\tilde{y}_2(x)/y_2(x)$ blows up in the limit $x\rightarrow \infty$ this implies that $\gamma>\beta$.

We will now construct the solution $\tilde{y}_2(x)$ for small $\epsilon$ by perturbing around the solution $y_2(x)$ i.e.
\begin{equation}
\tilde{y}_2(x)=y_2(x)+\epsilon z(x).
\end{equation}
To linear order in $\epsilon$ we find that $z(x)$ satisfies the ODE
\begin{equation}
z''(x)+(Q(x)+\lambda)z(x)=y_2(x),
\end{equation}
by means of variation of parameters we find that the relevant particular solution for $z(x)$ is given by
\begin{multline}
z(x)=y_2(x)\int_0^x y_1(x')y_2(x') dx' - y_1(x) \int_0^x y_2(x')^2 dx'\\ =e^{\beta x} p_2(x) \int_0^x p_1(x')p_2(x') dx' - e^{-\beta x} p_1(x) \int_0^x e^{2\beta x'}p_2^2(x')dx'.
\end{multline}
Due to the periodic nature of $p_2(x)$ it is bounded from above, take $M$ to be an upper bound such that for all $x$ $p_2(x)<M$ then
\begin{multline}
z(x)>e^{\beta x} p_2(x) \int_0^x p_1(x')p_2(x') dx' - e^{-\beta x} p_1(x) \int_0^x e^{2\beta x'}M^2dx'\\=e^{\beta x} p_2(x) \int_0^x p_1(x')p_2(x') dx' - e^{-\beta x}p_1(x)M^2\left( \frac{1}{2\beta} e^{2\beta x} -1\right)\\>e^{\beta x}\left( p_2(x)\int_0^x p_1(x')p_2(x') dx' - p_1(x)\frac{M^2}{2\beta}\right),
\end{multline}
The negative term in parenthesis is bounded whereas the positive term diverges linearly with $x$ in the limit $x\rightarrow \infty$, again due to the positive definite and periodic nature of $p_i(x)$. Looking back at the ratio $\tilde{y}_2(x)/y_2(x)$ in the large $x$ limit
$$
\lim_{x\rightarrow\infty} \frac{\tilde{y}_2(x)}{y_2(x)} = \lim_{x\rightarrow\infty} \frac{y_2(x)+\epsilon z(x) + O\left(\epsilon^2\right)}{y_2(x)}=\lim_{x\rightarrow\infty} 1+ \epsilon \left(\int_0^x p_1(x')p_2(x') dx' - \frac{M^2}{2\beta}\frac{p_1(x)}{p_2(x)}\right) + O(\epsilon^2),
$$
this ratio diverges to positive infinity in the limit $x\rightarrow \infty$, from which we can establish that $\gamma>\beta$. This completes the proof that the Floquet factor is a stricly decreasing function of $\lambda$ as long as $\lambda<\lambda_0$.

\FloatBarrier

\end{document}